\documentclass[aip, preprint]{revtex4-2}

\usepackage{graphicx}

\usepackage{algorithm}
\usepackage{algpseudocode}
\usepackage{setspace}
\usepackage{multirow}

\draft 

\begin{document}

\title[Machine Learning Stopping Power]
{ESPNN: a novel Electronic Stopping Power neural-network code built on 
the IAEA stopping power database. I. Atomic targets} 
\author{F. Bivort Haiek} 
\affiliation{Miner\'{\i}a de Datos y
Descubrimiento del Conocimiento, Universidad de Buenos Aires, Argentina.}
\author{A.M.P. Mendez} 
\affiliation{Instituto de Astronom\'{\i}a y F\'{\i}sica del Espacio, 
CONICET and Universidad de Buenos Aires, Argentina.}
\author{C.C. Montanari} 
\affiliation{Instituto de Astronom\'{\i}a y F\'{\i}sica del Espacio, 
CONICET and Universidad de Buenos Aires, Argentina.}
\author{D.M. Mitnik} 
\email{dmitnik@df.uba.ar}
\affiliation{Instituto de Astronom\'{\i}a y F\'{\i}sica del Espacio, 
CONICET and Universidad de Buenos Aires, Argentina.}

\makeatletter
\def\@email#1#2{%
 \endgroup
 \patchcmd{\titleblock@produce}
  {\frontmatter@RRAPformat}
  {\frontmatter@RRAPformat{\produce@RRAP{*#1\href{mailto:#2}{#2}}}\frontmatter@RRAPformat}
  {}{}
}%
\makeatother

\date{\today}

\begin{abstract}

The International Atomic Energy Agency (IAEA) stopping power database is 
a highly valued public resource compiling most of the experimental 
measurements published over nearly a century. The database--accessible 
to the global scientific community--is continuously updated and has 
been extensively employed in theoretical and experimental research for 
more than thirty years. 
This work aims to employ machine learning algorithms on the 2021 
IAEA database to predict accurate electronic stopping power cross 
sections for any ion and target combination in a wide range of 
incident energies.
Unsupervised machine learning methods are applied to clean the database 
in an automated manner. These techniques purge the data by removing 
suspicious outliers and old isolated values. A large portion of the 
remaining data is used to train a deep neural 
network, while the rest is set aside, constituting the test set.
The present work considers collisional systems only with atomic  
targets. 
The first version of the {\sc espnn} (electronic stopping power 
neural-network code), openly available to users, is shown to yield predicted 
values in excellent agreement with the experimental results of the test set.
\end{abstract}


\maketitle
\section{Introduction}
\label{sec:Introduction} 

At the end of 2015, the Nuclear Data Section of the International Atomic 
Energy Agency~\cite{iaea}  (IAEA) 
inherited the monumental work done by Paul~\cite{Paul:91, Paul:03,
Paul:13, Paul:13b, Paul:15, Paul:15a}. He collected about 1000 
experimental stopping power measurements made in multiple laboratories 
worldwide, including publications from as early as 1928. He kick-started 
his database project in 1990 at the University of Linz, and it has been 
available to the scientific community since then. The IAEA assumed the 
responsibility of maintaining, updating, and disseminating this data 
collection~\cite{Paul:15}, which included tables, figures, and 
comparisons of the published stopping data for ions in atomic targets, 
compounds, and new materials of technological interest. An overview of 
the database contents can be found in the review of Montanari and 
Dimitriou~\cite{Montanari:17}.

The stopping power is the mean energy loss per unit path length of the 
projectile in many collisional processes. Calculating the electronic 
stopping power involves determining the target system probabilities of 
occupying any electronic state different from the initial one due to the 
transfer of energy from the ion to the target electrons. Several reviews 
on this subject are available in the literature~\cite{Sigmund:06, 
Sigmund:14, Sigmund:16}. Different 
methods and semiempirical codes freely available online are linked and 
scrutinized on the IAEA stopping power website~\cite{iaeaprograms}.
Certainly, the code most widely used is {\sc srim}. This code is based 
on a semiempirical method developed by Ziegler~\cite{Ziegler:85}.
As reported in their work~\cite{Ziegler:10}, it reproduces 64\% of the 
data with an overall accuracy of 5\%. Noteworthy, the latest version of 
{\sc srim} includes measurements only up until 2013. 
Essential differences between the {\sc srim} predictions and new 
measurements have been reported since 
then~\cite{BA:15, Mat:21, Mor:20,Sel:20}.

The present work is the first of a series of publications where we design a 
robust and general model to accurately calculate the electronic stopping 
power in different target materials and along an extended energy range. 
To fulfill this task, we developed a machine learning (ML) model based 
on a clustering technique and a deep neural-network (NN) method.

\begin{figure}[H]
  \centering
  \includegraphics[width=0.9\textwidth]{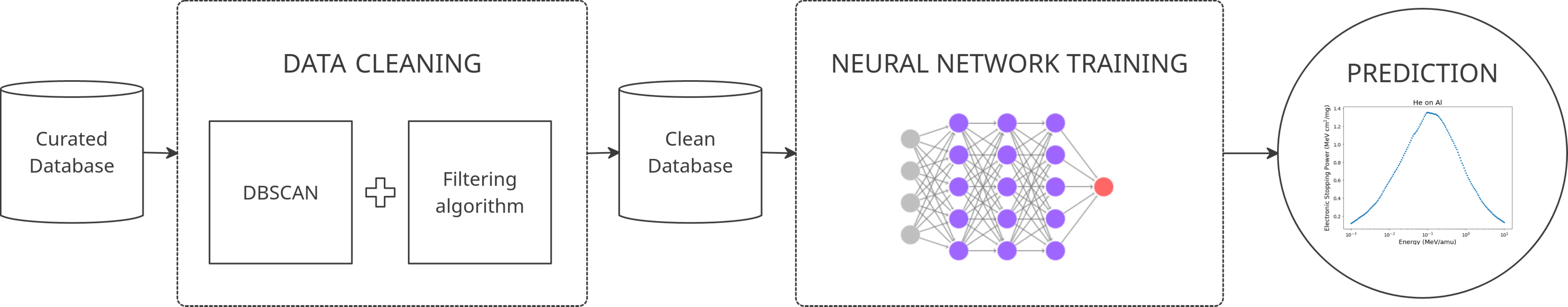}
  \caption{Graphic scheme of the electronic stopping power with 
  neural-network ({\sc espnn}) model. See the text for details.}
  \label{fig:espnn-model}
\end{figure}

The IAEA database was built by gathering published articles from diverse  
authors; hence, it needs to be standardized in its original form. 
The data are presented in various units and formats. Depending on the 
ion--target system, the experimental values are given as stopping power 
per unit length or cross sections per mass or atom. 
The database contains dozens of thousands of input values only for 
mono-elemental targets.
As a preliminary work, we devoted significant efforts to reorganizing 
the database, unifying the units, and arranging the data in a standard 
(csv) format, which enabled easy and quick access to the compiled data.
The first task consisted of cleaning the curated data 
outside the general trend. Purging these data by hand requires 
considerable amounts of work and is not recommended. 
Instead, we developed an unsupervised-machine-learning-based 
method to clean up the 
database by implementing a filtering algorithm and a cluster analysis. 
This clustering technique, called {\sc dbscan}, identifies outlier 
values and determines which data to keep in the cases of inconsistent 
overlapping. The cleaning procedure is shown schematically in the left 
dashed-box of Fig.~\ref{fig:espnn-model}.

As illustrated in Fig.~\ref{fig:espnn-model}, the cleaned database 
becomes the input of the second supervised-machine-learning method 
consisting of a deep neural-network. 
This network has many basic units (neurons) arranged in layers.  
Each neuron receives input signals from the previous elements, 
processes it through some weighted non-linear function, and transmits 
the resulting output to neurons belonging to the next layer. 
The neural-network contains a known input 
(projectiles, targets, energies), and a known output 
(the corresponding experimental stopping power). 
The training is performed by adjusting the 
weights, minimizing the differences between the final processed output 
of the network (predictions) and the experimental results.
In that way, the model can  accurately reproduce the experimental 
values and, hopefully, predict new results in the 
cases not included in the training procedure (the test set). 
The resulting model and code {\sc espnn} 
(electronic stopping power with neural-network) are presented in this work. 
In this first article, we report the results obtained only for atomic
targets. 
Our results were obtained with excellent accuracy using different error 
metrics, such as MAPE (Mean Average Percentage Error) or MAE (Mean 
Absolute Error).

Section~\ref{sec:cleansing} describes the machine learning (ML) methods 
employed to depurate the database, showing a few examples of the 
original data and the remaining final input employed in the network training. 
In Section~\ref{sec:ML}, the deep neural-network architecture is 
described. An overview of the training, validation and test sets is 
presented; some details concerning the training procedure are also 
discussed. Section \ref{sec:results} presents selected results and the
analysis of distinct error metrics, which give a sense of the method's 
accuracy. In Appendix~\ref{app:code}, we provide instructions for 
installing and using the {\sc espnn} code.

\section{Database cleansing}
\label{sec:cleansing}

\subsection{Database review}
\label{subsec:review}

The database (updated in December 2021) consists of 60173 experimental 
measurements, representing stopping power values for 1491 ion--target 
combinations of 49 projectiles colliding with 283 targets across the 
energy range $10^{-4}-10^4$~MeV/amu and ion and target atomic 
masses from 1 to 240 amu. 
Concerning only the mono-elemental targets, there are 706 collision cases  
composed of 44 different projectiles and 73 targets, resulting in 
36544 experimental data points. The experimental data summarize 1190 
publications covering the period 1928--2021. The reorganization of the 
database allows for performing extensive statistical analysis. This task 
would indicate, for instance, the lack of data in certain energy regions, 
over-measured and under-measured systems, which may guide experimental 
groups regarding the necessity of new findings. A detailed analysis of 
the database's current status will be presented in a forthcoming article.

The raw data collected come from several publications; the results of 
the same ion--target collision may show significant discrepancies (much larger than the 
error of the individual set of experimental data). Cleaning the database 
is crucial; well-thought criteria must be adopted for selecting the most 
reliable values, accompanied by careful examination of the outcome.
The immense difficulties of scrutinizing such an extensive dataset are 
resolved by implementing a straightforward ML-based method. First, the 
{\sc dbscan} classification algorithm is used to group similar results 
in clusters and to identify outliers, i.e., values suspected of being 
erroneous. Then, an algorithm is developed to assess clusters and 
outliers by introducing different criteria for overlapping and isolated 
data. In the following, we briefly explain these algorithms.

\subsection{The {\sc dbscan} algorithm}
\label{subsec:dbscan}

Clustering algorithms are attractive for the task of class identification 
in spatial databases. However, most well-known unsupervised classification
algorithms suffer severe drawbacks when applied to large spatial databases. 
That is, elements in the same cluster may not share enough similarities, 
or performance may be poor. Also, while partition-based algorithms, such 
as K-means, may be easy to understand and implement in practice, the 
algorithm has no notion of outliers; all points are assigned to a cluster, 
even if they do not belong to any. Moreover, anomalous points draw the 
cluster's centroid toward them, making it more difficult to classify 
them as anomalous points.
In contrast, density-based clustering locates regions of high density 
that are separated by regions of low density (in this context, density 
is defined as the number of points within a specified radius).

In this work, we used the density-based spatial clustering of applications 
with noise algorithm {\sc dbscan}~\cite{Ester:96,Schubert:17}. 
The main idea behind this technique is the following: 
for a set of points in some space, the 
algorithm groups together values that are closely packed (points with 
many nearby neighbors), marking as outlier points that lie alone in 
low-density regions (whose nearest neighbors are too far away). 
It requires two input parameters: the radius of the neighborhood, 
$\epsilon$, and the number of reachable points (within a distance 
$\epsilon$), $N_{\mathrm{min}}$, required to form a dense region.
All points belonging to an $\epsilon$-neighborhood configure a cluster.
All points not reachable from any other point are outliers or noise points.
{\sc dbscan} is significantly effective in discovering clusters of 
arbitrary shapes, which makes it a standard clustering algorithm and one 
of the most cited in the scientific literature. This method has numerous 
advantages: it does not require one to specify the number of clusters in 
the data {\it a priori} (as opposed to K-means, for example).
{\sc dbscan} can find arbitrarily shaped clusters; it has a notion of 
noise, is robust to outliers, and is mainly insensitive to the data 
order.

        \begin{figure}[H]
            \centering
             \includegraphics[width=0.375\textwidth]{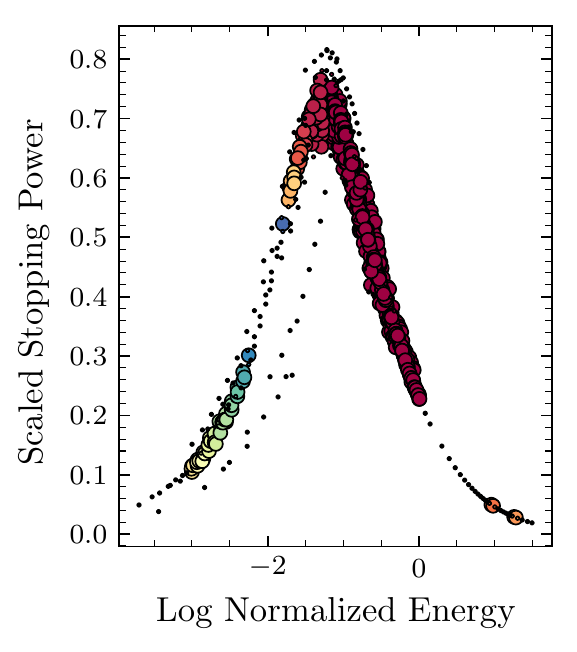}
             \includegraphics[width=0.375\textwidth]{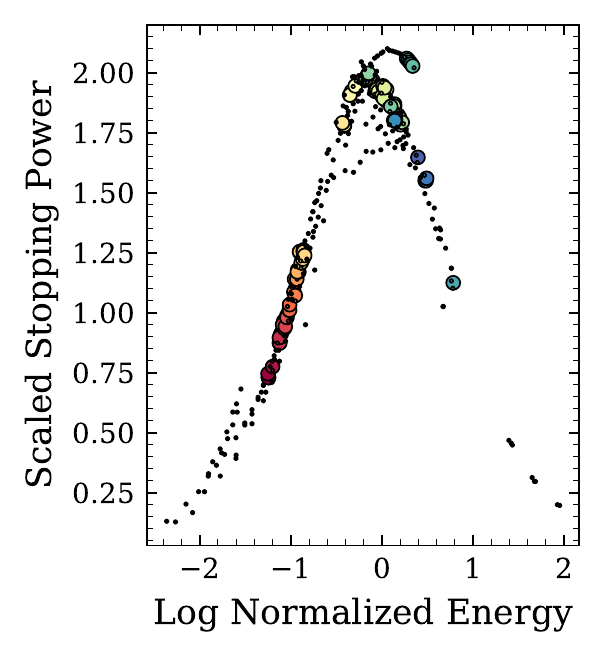}
             \caption{Scaled stopping power cross sections 
                       for H in Si (left) and O in Au (right). 
                       The different colors represent clusters detected 
                       by the {\sc dbscan} algorithm with 
                       $\epsilon=0.025$ and $N_{\mathrm{min}}=3$. 
                       Outliers points are indicated with small dots. }
        \label{fig:dbscan} 
        \end{figure}    

As examples of the clustering partitions produced by {\sc dbscan},
we show in Fig.~\ref{fig:dbscan} the stopping power cross sections 
for H in Si (left), and for O in Au (right). In all cases, the 
cross sections have been scaled to keep both axes of similar size. 
We found it convenient to set the input parameters as $\epsilon=0.025$ 
and $N_{\mathrm{min}}=3$. 
For H in Si, the unsupervised algorithm detected 19 clusters (each one 
plotted with a different color). The small black dots without 
$N_{\mathrm{min}}$ neighboring points inside a circle of a radius 
$\epsilon$ are considered noise points. In total, the algorithm labeled 
139 outliers. For O in Au, 26 clusters and 181 outliers were identified.

\subsection{Filtering procedure}
\label{subsec:filter}

When the {\sc dbscan} algorithm is used for data cleansing, the assigned 
outliers are generally and systematically removed from the dataset. 
However, the amount of data to be cleaned by following this standard 
course of action would be significant. Instead, we designed and 
implemented a three-step sequential filtering algorithm based on the 
clustering results. 

Once the clusters and the outliers are identified, the sequential 
algorithm determines which data points are kept. First, we inspect 
energy regions with single and multiple publications based on 
a year-of-publication criterion. 
For example, in Fig.~\ref{fig:dbscan}, the data points marked as 
outliers by {\sc dbscan} may cover energy regions with no other 
available values. This context is essential and is considered by our 
cleaning algorithm: experimental values belonging to an energy region 
with no other measurements are considered valid data points; diversely, 
they are examined by the date criterion. 
In the following step, we examine overlapping (in energy) but isolated 
(in cross section) data points. 
Only those fulfilling an {\it ad-hoc} criterion are kept; 
otherwise, they are tagged as true outliers. Finally, we inspect 
clustering composition, i.e., whether a cluster comprises data from 
single or many publications. 
The general structure of this cleaning procedure is described 
in the following, while a pseudocode scheme resuming the filtering 
algorithm is presented in Appendix~\ref{app:filtering}.

For each collisional system, every publication is first assessed 
according to the energy overlapping with other references. We determine 
the number of experimental values $n_i$ published in the publication
$P_i$, which are spread over an energy range $\Delta E_i$. 
If the publication data cover a region with no other measurements, these 
data are included in the input database. Otherwise, a date  
criterion is followed; for the remaining publications of the same 
collisional system, we count the experimental values that: 
1) fall inside the energy range $\Delta E_i$ and 
2) have been published after $P_i$. 
The total energy range covered by these posterior publications  
is denoted as $\Delta E_i^p$. We evaluate what portion of the energy 
range covered by the publication $P_i$ is also covered by newer 
publications; i.e., 
\begin{eqnarray}
  \mathrm{If} ~~ \frac{\Delta E_i^p} {\Delta E_i} \leq \sigma_{\Delta} ~~ 
  \mathrm{\to \ {\bf KEEP} ~~the ~publication } ~P_i   \, .
  \label{eq:condDelta}
\end{eqnarray}
The chosen overlap parameter $\sigma_{\Delta}$ is close to 0.6 except 
for the cases in which the number of references addressing the specific 
collisional system is tiny; in such cases, $\sigma_{\Delta} \approx 1$.
The condition in (\ref{eq:condDelta}) ensures that if a publication 
presents results in an unexplored energy range,  
these values are kept and included in the input of the neural-network, 
no matter if they belong to a cluster or are classified outliers.  
An example of this criterion is shown in the right panel of 
Fig.~\ref{fig:dbscan}; the experimental low energy values for O in Au 
are kept in the input database even though {\sc dbscan} considers them 
as outliers.

Next, we deal with the remaining energy-overlapping results. For a given 
ion--target system, the second filtering step consists of dropping 
``isolated'' results, i.e., data points appearing at the same energy 
range as others but having different values. In this instance, we count 
the number of outliers $N^i_{\mathrm{outl}}$ detected in $P_i$ and
evaluate
\begin{eqnarray}
\mathrm{If} ~~ \frac{N^i_{\mathrm{outl}}}{n_i} > \sigma_{\mathrm{out}} ~~ 
\mathrm{{\to \ \bf DROP} ~~the ~publication } ~P_i
\label{eq:condOutliers}
\end{eqnarray}
where $\sigma_{\mathrm{out}} \approx 0.45$, except for when the number 
of publications is small; then, $\sigma_{\mathrm{out}} \approx 1$.
This criterion implies that if a publication has many results 
overlapping in energy with newer measurements, but most of their values 
are considered suspiciously wrong, it is convenient to keep the newer 
values rather than the outliers.

Finally, we perform the same test for isolated results, but instead of 
inspecting the outliers, we look into the clusters. We identify the 
largest cluster $C_i$ in a particular publication $P_i$. 
This cluster has a total of $t^i$ data points, including the 
$l^i$ values belonging only to $P_i$.
We aim to verify if the biggest cluster of a publication is formed 
by results obtained from many sources and not just from this unique work.
Thus, we evaluate the condition
\begin{eqnarray}
\mathrm{If} ~~ \frac{l^i}{t^i} > \sigma_{\mathrm{clu}} ~~ 
\mathrm{{\to \ \bf DROP} ~~the ~publication } ~P_i
\label{eq:condClusters}
\end{eqnarray}
with $\sigma_{\mathrm{clu}} \approx 0.45$, except for cases where the 
number of publications is small; in such cases 
$\sigma_{\mathrm{clu}} \approx 1$.

\begin{figure}[H]
        \centering
        \includegraphics[width=0.80\textwidth]{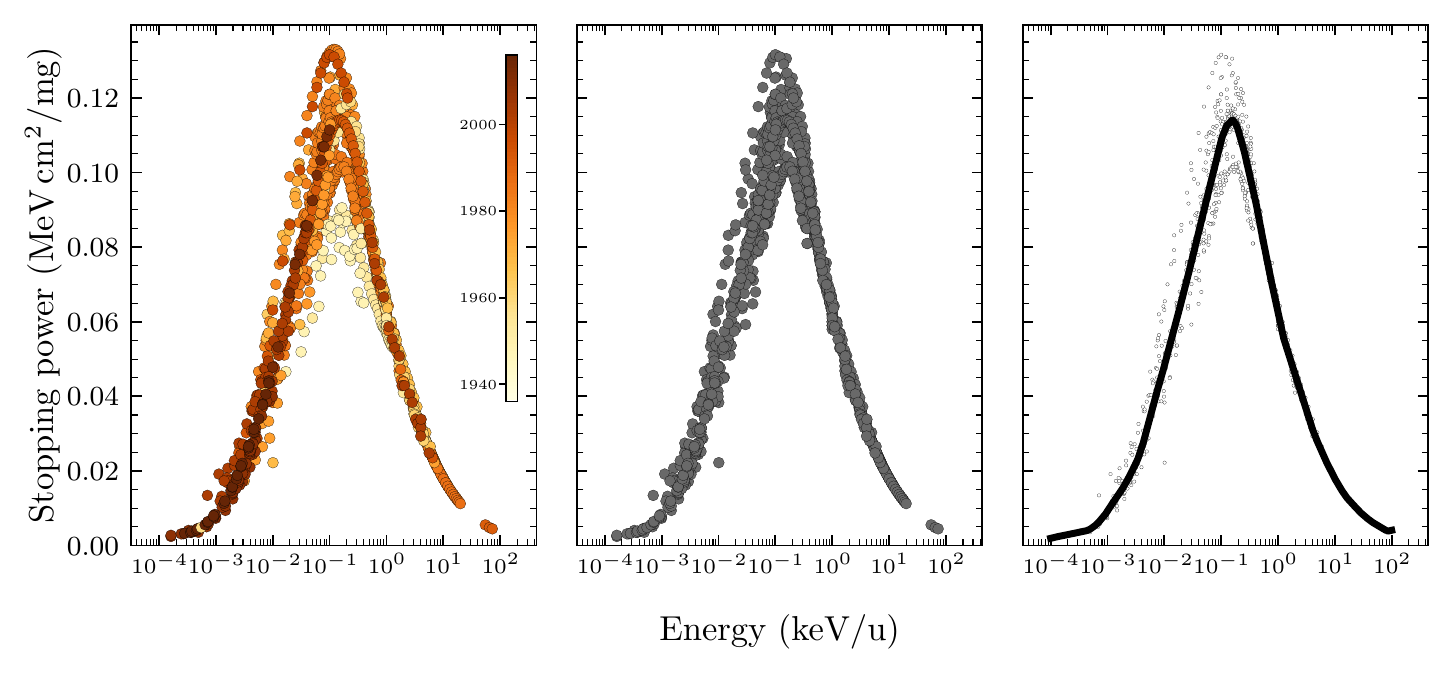}\\
        \includegraphics[width=0.80\textwidth]{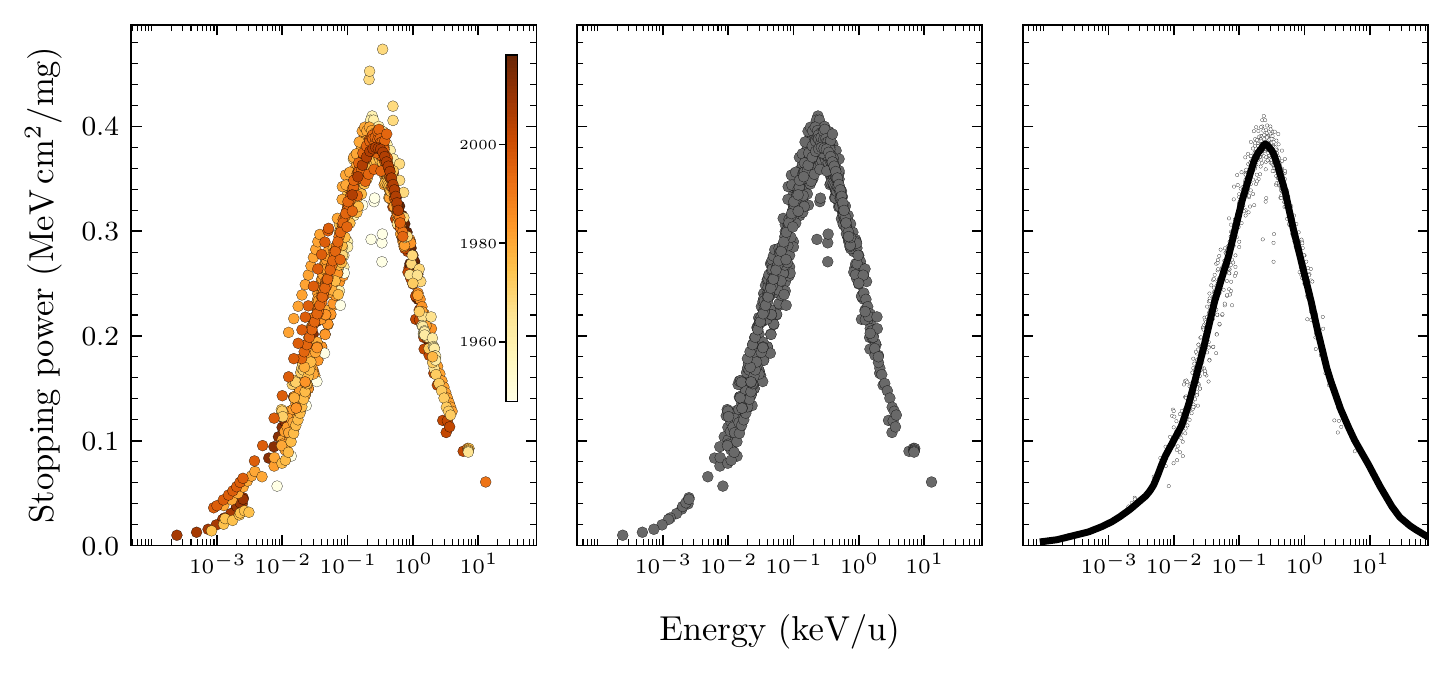}\\
        \includegraphics[width=0.80\textwidth]{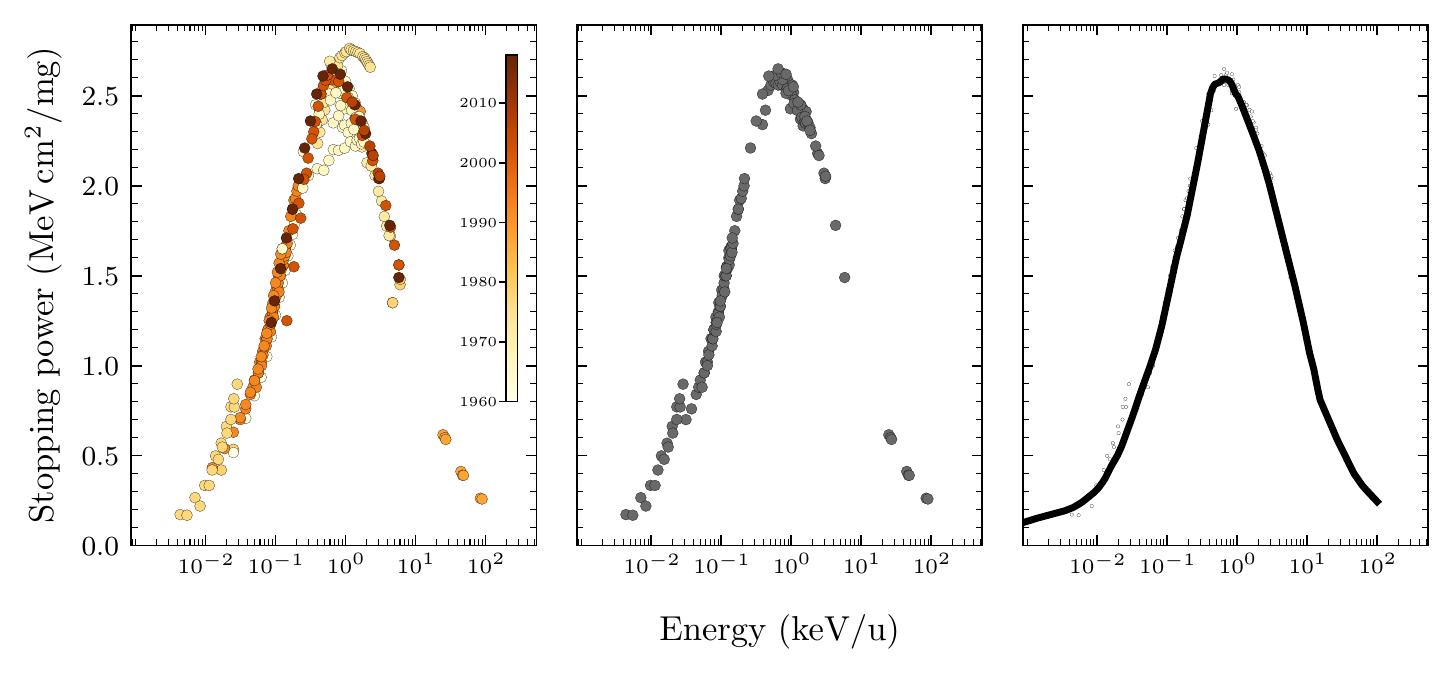}\\
        \caption{
            Left: Experimental results for stopping power cross sections 
            of collisional systems with Au as targets. 
            The colors indicate the year of publication of the data. 
            Center: Filtered data, resulting from the cleaning
            procedure explained in the text.
            Right: Predicted data from the neural-network.
            Top: H in Au. 
            Middle: He in Au. 
            Bottom: O in Au.}
        \label{fig:filtradosAu} 
        \end{figure}

\begin{figure}[H]
        \centering
        \includegraphics[width=0.80\textwidth]{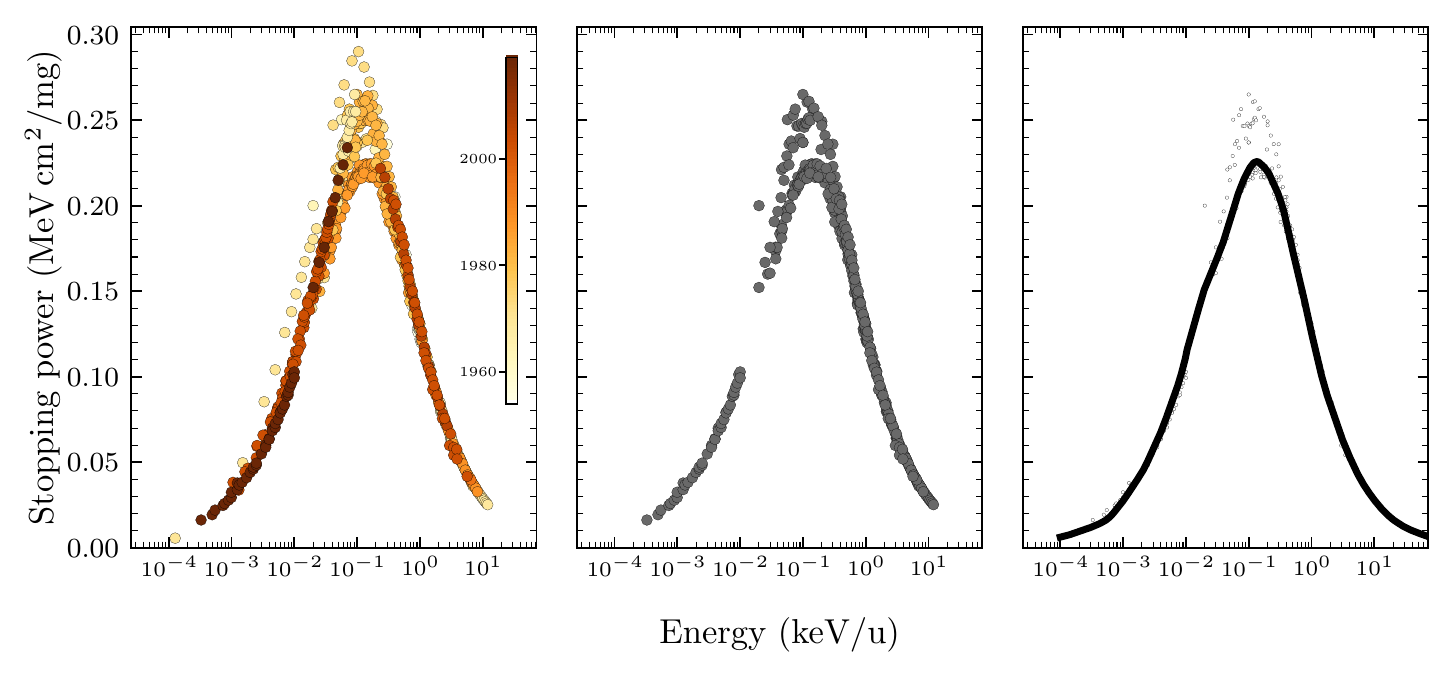}\\
        \includegraphics[width=0.80\textwidth]{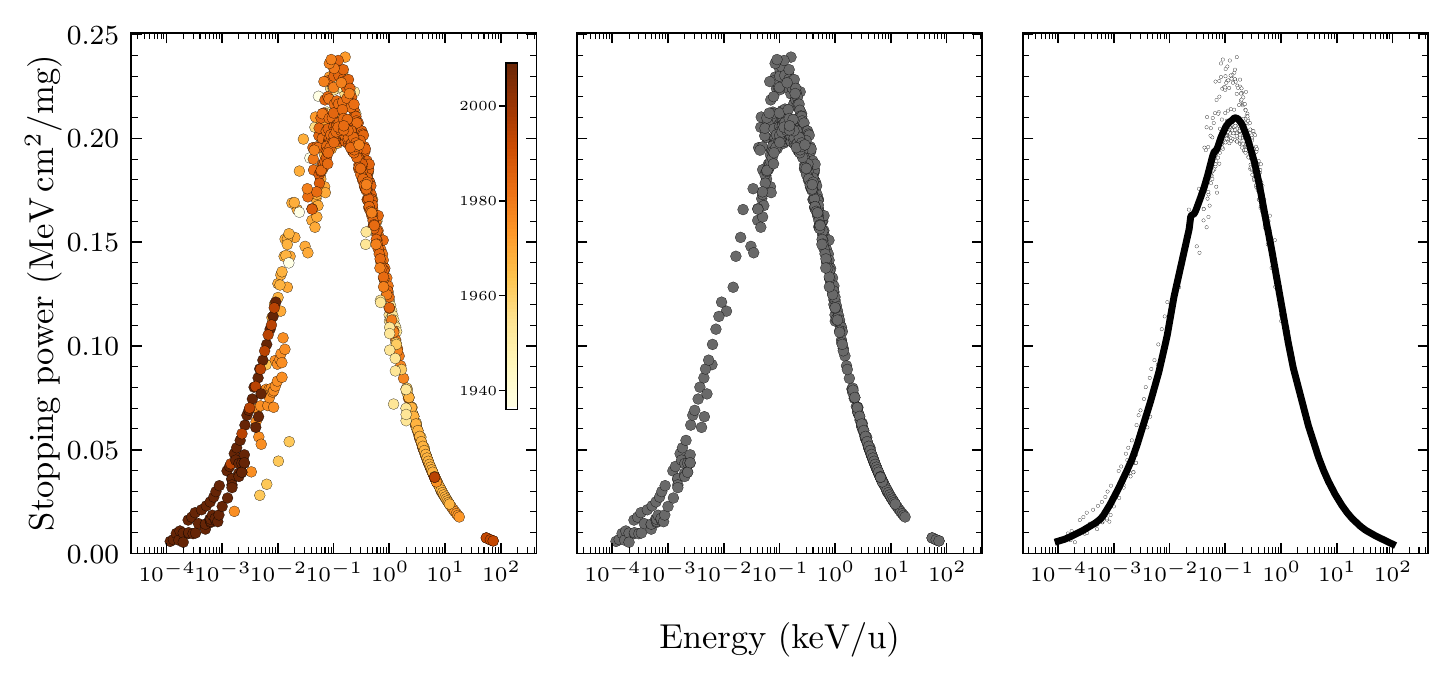}\\
        \includegraphics[width=0.80\textwidth]{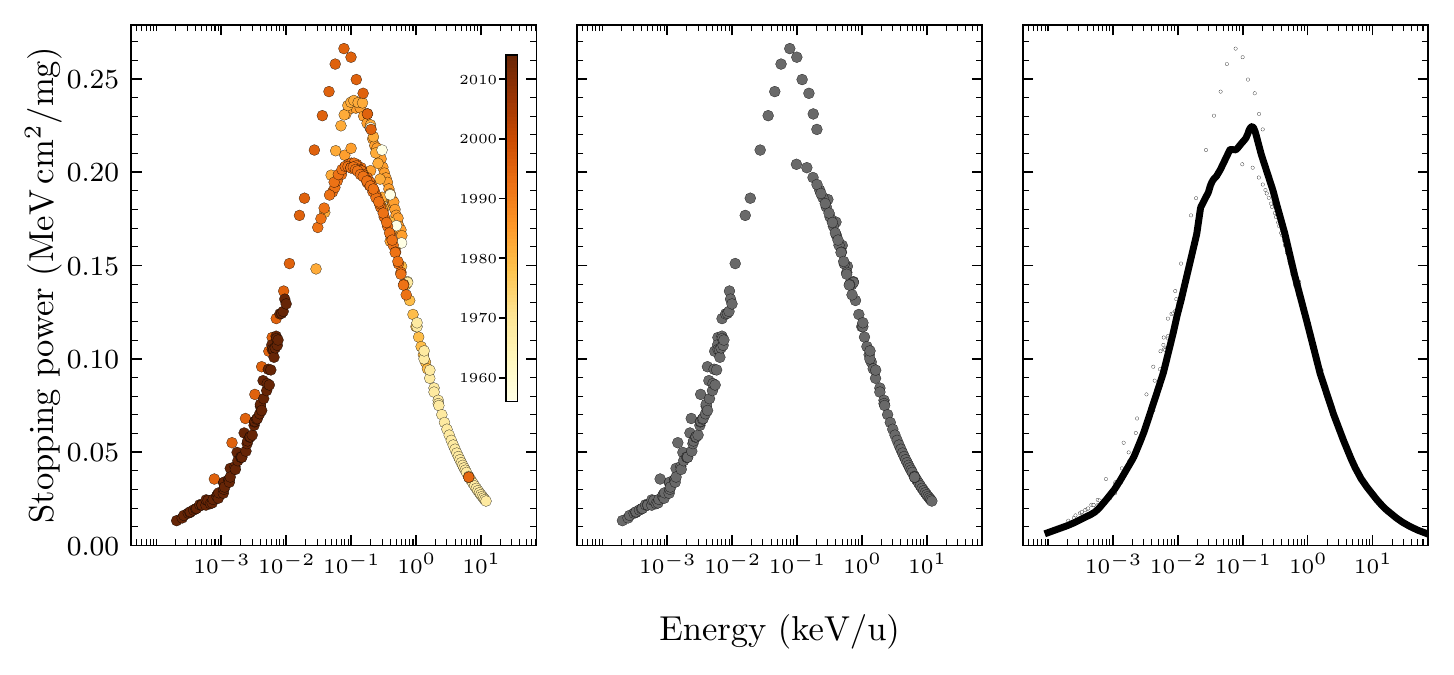}\\
        \caption{
            Left: Experimental results for 
            stopping power cross sections, for H projectiles. 
            The colors indicate the year of publication of the data. 
            Center: Filtered data.
            Right: Predicted data from the neural-network.
            Top: H in Ni. 
            Middle: H in Cu. 
            Bottom: H in Zn.}
        \label{fig:filtradosH} 
        \end{figure}

\begin{figure}[H]
        \centering
        \includegraphics[width=0.80\textwidth]{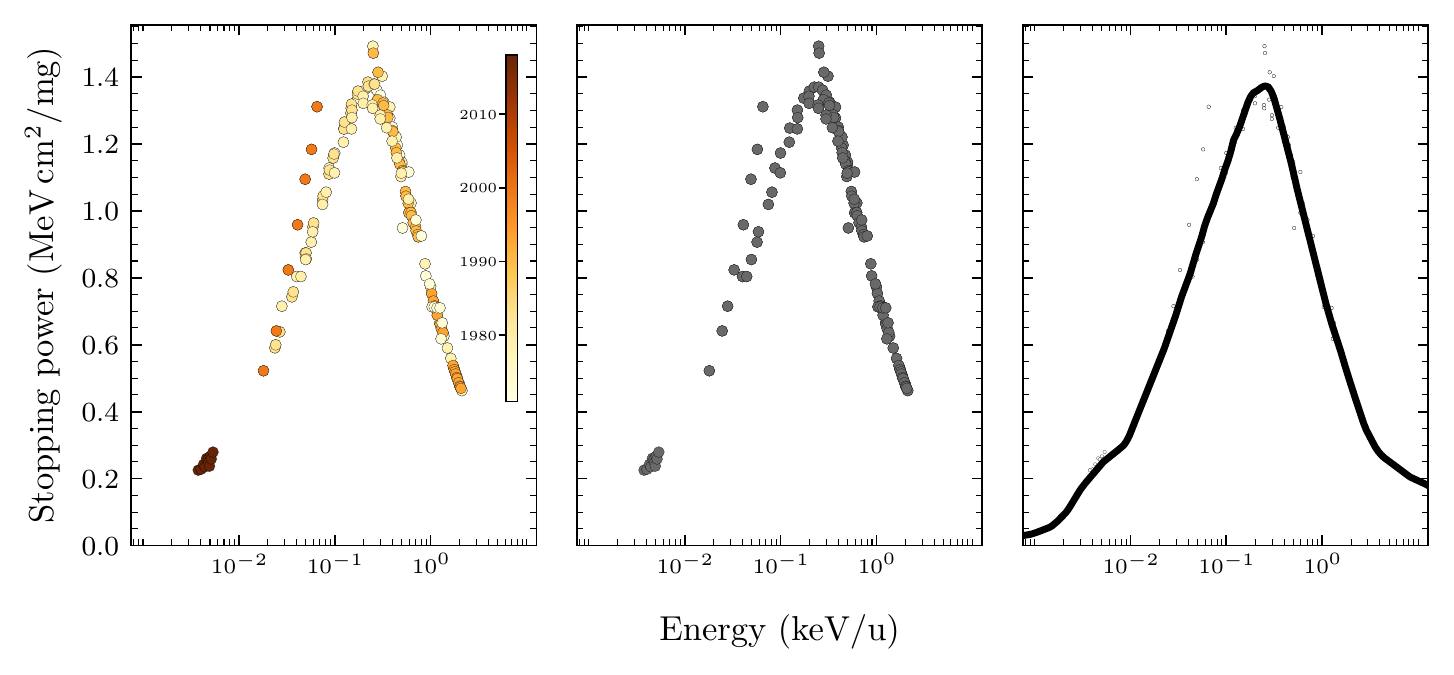}\\
        \includegraphics[width=0.80\textwidth]{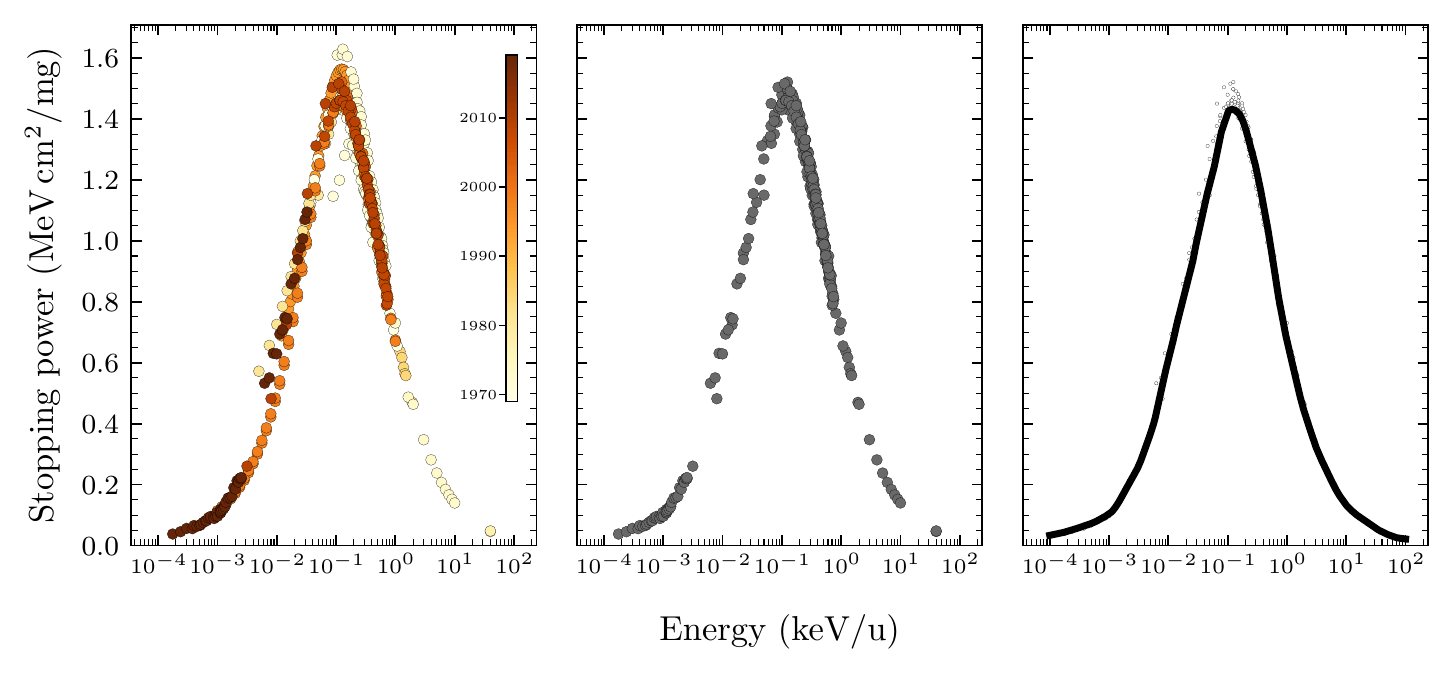}\\
        \includegraphics[width=0.80\textwidth]{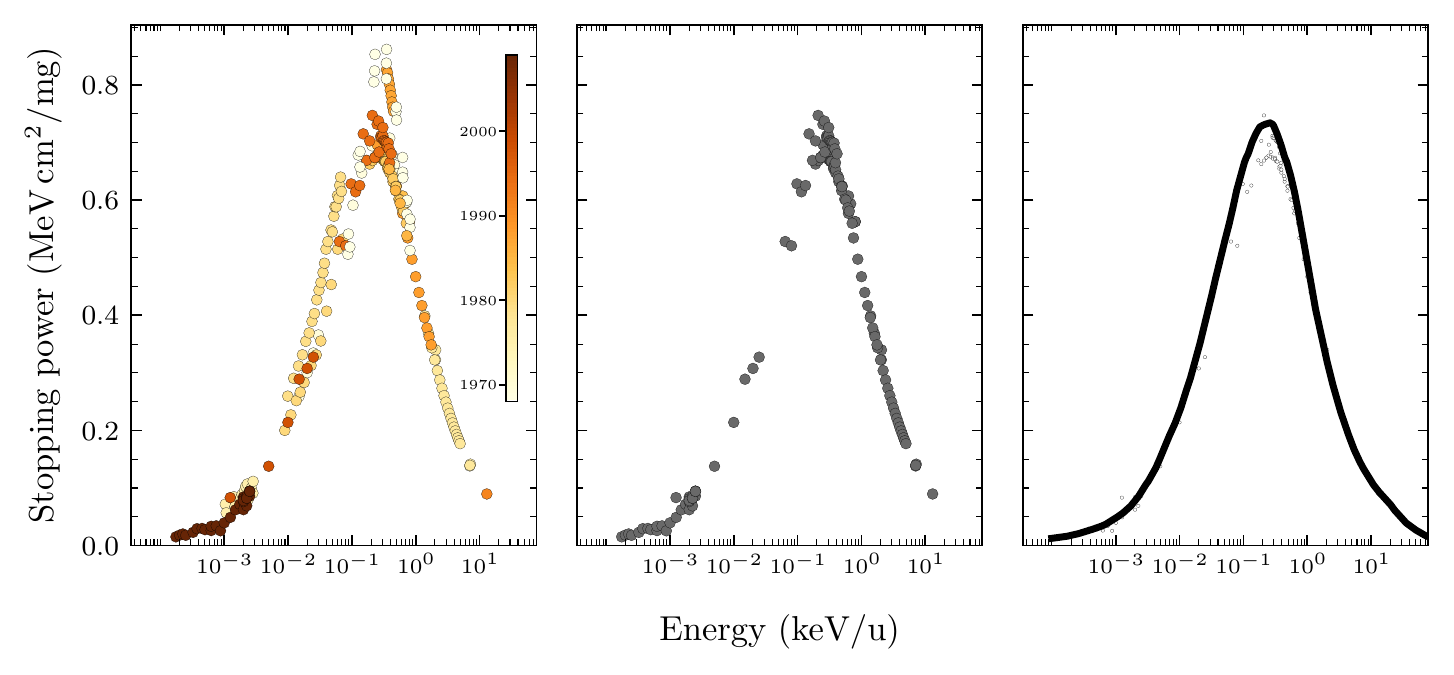}\\
        \caption{
        Left: Experimental results for 
        stopping power cross sections, for He projectiles. 
        The colors indicate the year of publication of the data.
        Center: Filtered data.
        Right: Predicted data from the neural-network.
        Top: He in Ne. 
        Middle: He  in Si. 
        Bottom: He in Cu.}
        \label{fig:filtradosHe} 
        \end{figure}

\begin{figure}[H]
        \centering
        \includegraphics[width=0.80\textwidth]{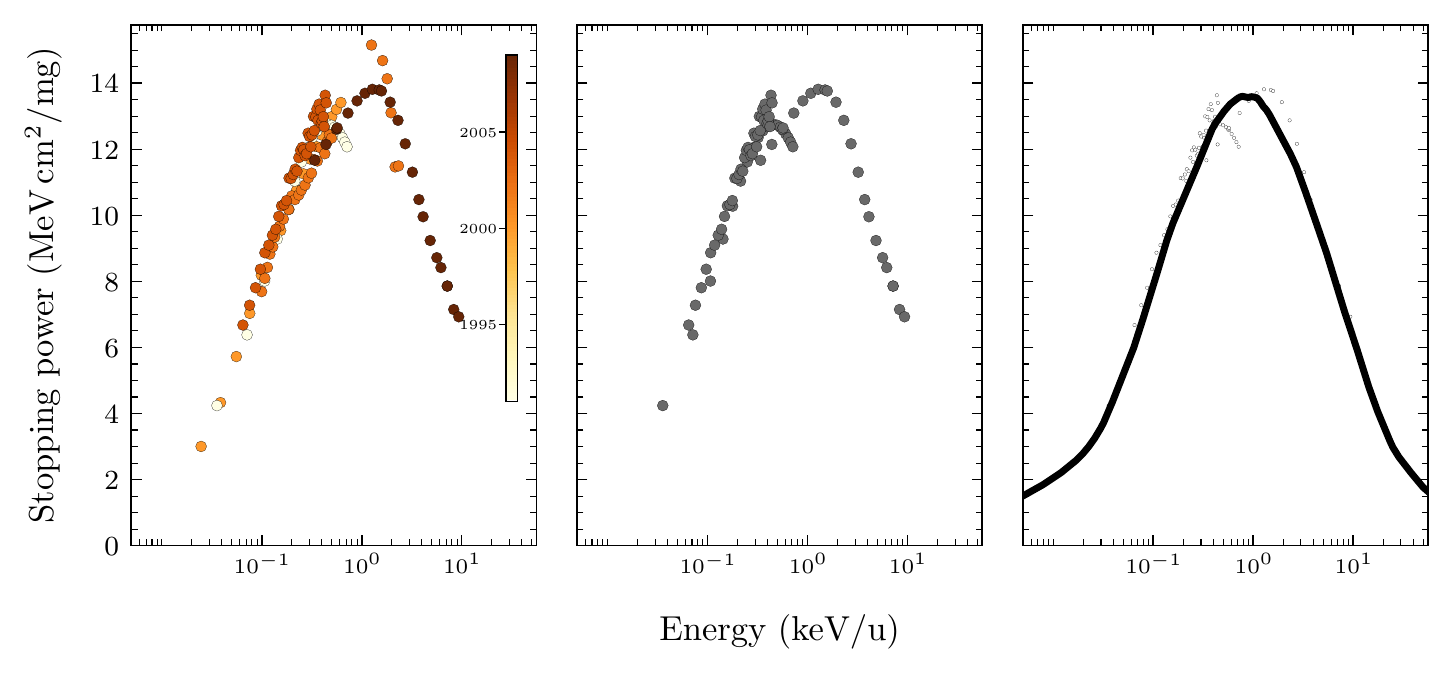}\\
        \includegraphics[width=0.80\textwidth]{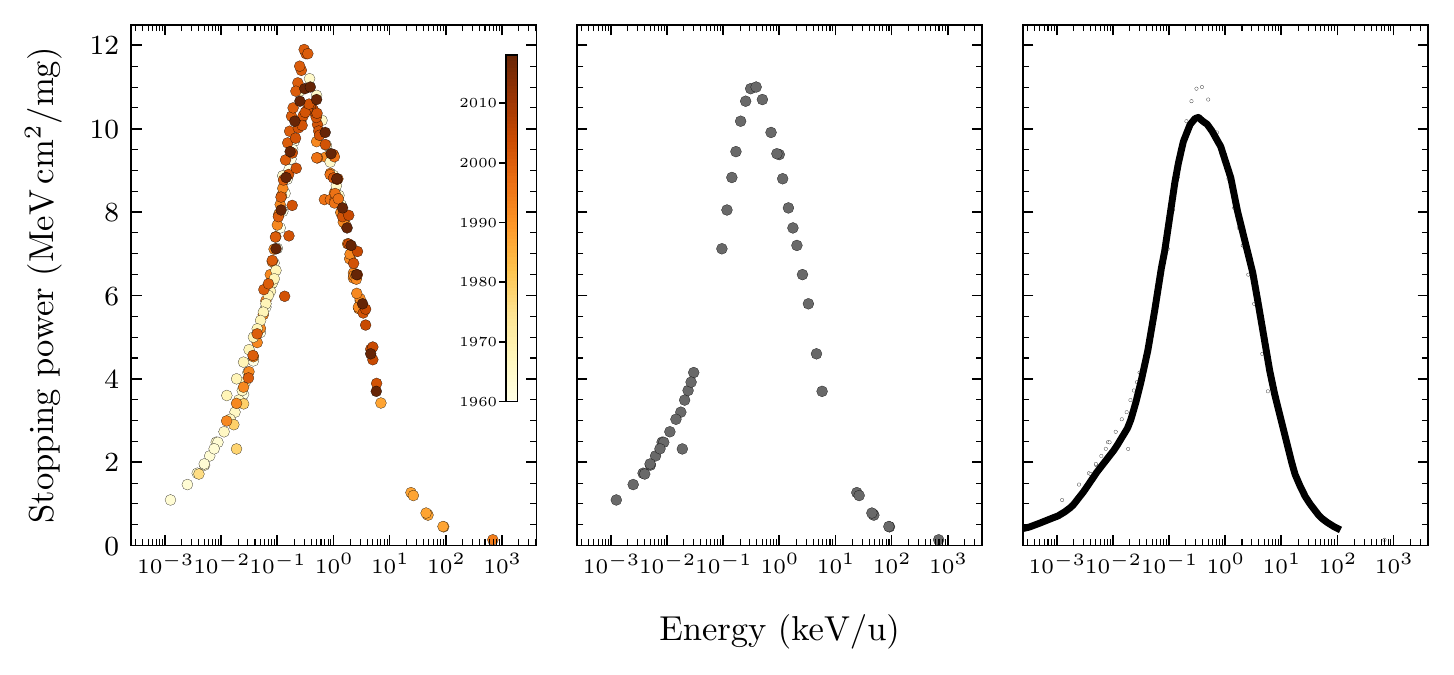}\\
        \includegraphics[width=0.80\textwidth]{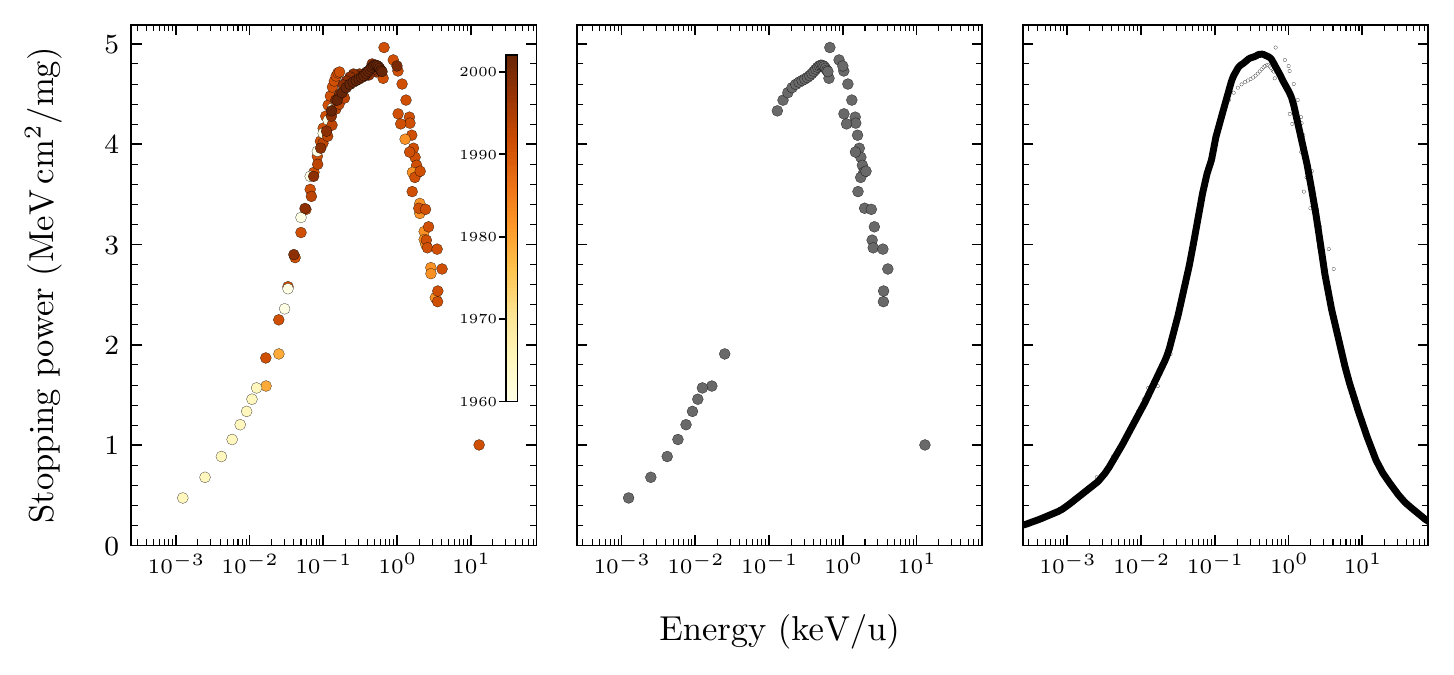}\\
        \caption{
        Left: Experimental results for 
        stopping power cross sections. 
        The colors indicate the year of publication of the data.
        Middle: Filtered data.
        Right: Predicted data from the neural-network.
        Top: Si in Si. 
        Middle: O in C. 
        Bottom: C in Al.}
        \label{fig:filtradospred} 
        \end{figure}

The performance of the cleaning procedure is illustrated with many 
examples in Figs.~\ref{fig:filtradosAu}, \ref{fig:filtradosH}, 
\ref{fig:filtradosHe}, and \ref{fig:filtradospred}. The left panels of 
all these figures display the complete data set for each collisional 
system. The central panels illustrate the cleaned data set, i.e., the 
outcome of implementing the cleaning procedure. 

In Fig.~\ref{fig:filtradosH}, we show the electronic stopping power 
cross section of projectiles H (top row), He (middle row), and O 
(bottom row), colliding with Au. For this target, a large amount of data 
is available. However, for these collisional systems, significant 
discrepancies among the various experimental results are evident, 
stressing the importance of an effective filtering mechanism. For H in 
Au, the top-center subplot shows that the cleaning algorithm can safely 
discard only the data produced before 1980. The middle-center subplot 
indicates that many results have been dropped for He in Au, particularly 
in the low energy range. The left subplot for O in Au illustrates 
multiple clusters overlapping in the energy region close to the stopping 
power cross section maxima. In the bottom-center subplot, we observe 
that the algorithm chooses to keep the latest experiments, discarding 
the old data; as a result, a clean and smooth curve is obtained.

It is important to stress that the filtering algorithm favors the latest 
data, but that does not necessarily imply removing the older results. 
The procedure will not rule out older measurements if they are 
statistically valid. Of course, the data cleansing procedure can be more 
severe or relaxed, according to the parameters chosen. It is possible to 
change the {\sc dbscan} parameters to modify the size of the clusters 
and the filtering parameters from Eqs. (\ref{eq:condDelta}), 
(\ref{eq:condOutliers}), and (\ref{eq:condClusters}) for keeping or 
discarding some publications. Considering the success of the 
neural-network implementation that will be explored in the following, 
we assume that the values adopted are satisfactory for general purposes. 
An example of this feature is given in Fig.~\ref{fig:filtradosH}, where 
we display the electronic stopping power cross sections of H colliding 
with Ni (top row), Cu (middle row), and Zn (bottom row). In these cases, 
the filtering algorithm does not remove the (most) differences near the 
maxima. 

Fig.~\ref{fig:filtradosHe} shows stopping power cross sections of 
He colliding with Ne (top),  Si (middle), and Cu (bottom). In some cases, 
the filtering procedure (from the left to the center figures) is hardly 
noticeable; in others, the year-of-publication criterion in the cleaning 
algorithm plays an important role. Examples of these cases are He in Ne 
and He in Cu, respectively.

For projectiles heavier than He, the amount of experimental data in the 
database (and generally measured) is scarcer than for H and He. 
For example,  the cases Si in Si, O in C, and C in Al are displayed in 
Fig.~\ref{fig:filtradospred}. Although these collisional systems have 
fewer data points, the filtering procedure was performed without further 
consideration. The outcome in most cases is strongly noticeable, as the
center subplots from Fig.~\ref{fig:filtradospred} illustrate.

It was not apparent beforehand how severe the filtering had to be to 
proceed with the following supervised method. 
Hard filtering (resulting in clean and smooth data) seems to lead the 
neural-network to overfit the input values while allowing some ``noise'' 
helps to prevent this problem.
The present clustering-based algorithm reduced the original 36000 data 
points to 28000 values. 
These filtered measurements constitute the input 
database and are used to train a deep neural-network, which will 
be described in Sec.~\ref{sec:ML}.

\newpage

\section{Stopping Power predictions}
\label{sec:ML}

Calculating the stopping power of an ion when interacting with a target 
system involves numerous processes and parameters, which makes this a 
challenging problem. The different theoretical models developed to 
describe the experimental data accurately have not yet achieved 
total success. Semiempirical methods are generally needed to fit the 
known results and provide recommended values for multipurpose 
simulations and applications. Moreover, the experimental data also 
suffer from many discrepancies, requiring the introduction of auxiliary 
methods to classify and select the results. Considering the copious 
amount of experimental values in the IAEA database, it constitutes an 
ideal scenario where machine learning methods can help handle the 
complexities of this multiprocess phenomenon and even discover hidden 
relations in the data.

At the beginning of this project, two different papers following this 
approach were published. 
Parfitt and Jackman~\cite{Parfitt:20} trained a random forest regression
algorithm using thousands of measurements collected in Paul's stopping
power database (with data up to 2015). Based on extensive evaluations 
through $k$-fold cross-validation 
against several error metrics, they demonstrated that their model 
could fit the training data and make low-error predictions 
on unseen test data.
Using the same database, Guo {\it et al.}~\cite{Guo:22} trained a deep 
neural-network to reproduce the experimental results for elementary 
targets with good accuracy.

\subsection{Model}
\label{subsec:model}

While designing our neural-network, a thorough analysis was conducted to 
establish the best 
hyperparameters to use. This study includes not only the 
architectural design of the NN but also the definition of the loss 
function to minimize, the parameters involved in the minimization 
process, the learning rate, weight decay, and others. We found the best 
neural-network design consisting of a grid of five fully connected hidden 
layers, with $10 \times 24 \times 32 \times 24 \times 10$ elements, 
a leaky-ReLu activation function for every layer, and the adaptive 
moment estimation ({\sc Adam}) optimizer as the minimization 
algorithm. The dropout rates are 0.2 for the input and output layers and 
0.5 for every hidden layer. 
The other parameters for the training model are the 
learning rate $\alpha=0.001$, the batch size $b=64$, 
and the weight decay $\lambda=10^{-10}$. 
We employed 300 epochs with an 
early stopping step of 50. 
The re-parametrization trick \cite{Salimans:16} was used to increase 
the convergence speed.

It is worth mentioning the following technical detail. 
The stopping power must monotonically decrease towards low energies. 
However, for some collisional systems, an unphysical 
increase of the predicted values can appear toward small energy values. 
We tried adding minimal fictitious stopping power values at 
very low energies, but as expected, the training became extremely 
unstable due to its MAPE component. 
The correct behavior is accomplished in the final model by a simple trick: 
removing the bias parameters from the first linear layer.

An additional study focused on allocating the 
data for the learning process. 
From the cleaned data (28000 values), we separated 5\% (1400 results) 
for the test set. 
These data did not participate in the training process. 
The remaining 95\% of the data, called the learning set, was 
partitioned into five folds. Four folds comprised the training set, 
while the fifth fold constituted the validation set. Using this 
design, we trained the neural-network until the loss function 
successfully converged. Then, the weights of every layer are kept to 
estimate the stopping power cross sections in a subsequent stage. 
This procedure is carried out iteratively by rotating the 
training-validation folds. As a result, we obtain five sets of weights 
and, therefore, five different estimations for any input entry. 
The final predictions are attained by averaging the results obtained 
from the five NN weights, also providing a rough estimation of the 
uncertainty present in the calculation. 

As explained in Parfitt and Jackman~\cite{Parfitt:20}, 
different error metrics can be used to quantify the 
performance of the model in terms of the predicted values 
$y_{\mathrm{pred}}$ against the true experimental 
values $y_{\mathrm{true}}$. 
For comparisons with other methods, we use the 
mean absolute percentage error loss function, or MAPE, given by 
\begin{eqnarray}
\mathrm{MAPE} \equiv \frac{100}{n} \,\, \sum \, 
\left| 
\frac{ y_{\mathrm{true}} - y_{\mathrm{pred}} }{y_{\mathrm{true}} } \, 
\right|
\, .
\label{eq:mape}
\end{eqnarray}

\begin{figure}[H]
    \centering
    \includegraphics[width=0.75\textwidth]{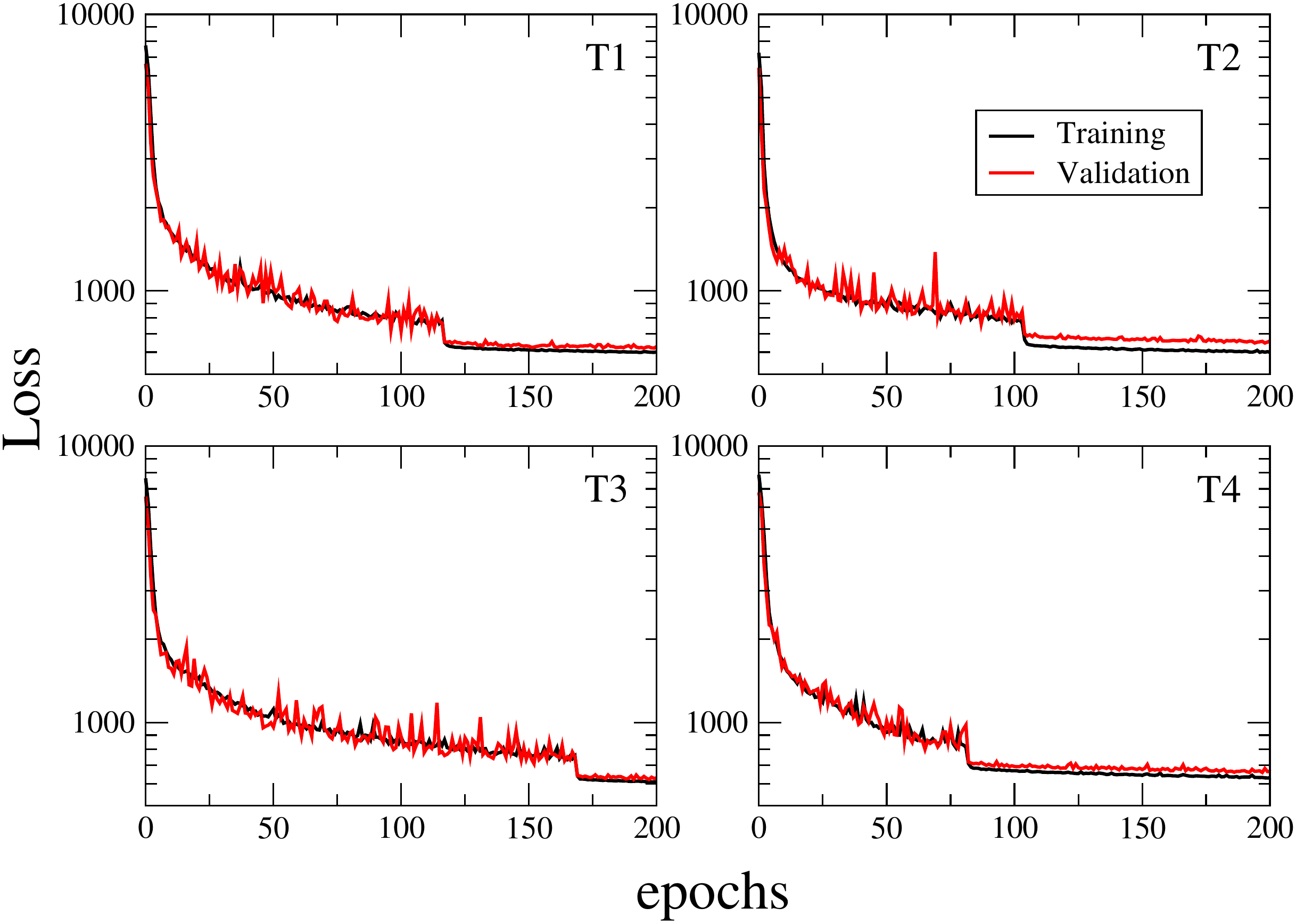} 
    \caption{Learning curve for the training and  
    validation datasets, for four different folds.
    \label{fig:training} }
\end{figure}

However, for the training process, the loss function is better 
defined by adding a small contribution of another metric to the MAPE. 
This prevents the training from featuring instabilities, which can 
appear at small stopping power values. 
We defined the loss function as a linear combination of two distinct 
metrics,
\begin{eqnarray}
L \equiv  \mathrm{MAPE} + \frac{\mathrm{MSE}}{\beta}  \,\, ,
\end{eqnarray}
where MSE is the mean-squared error defined by 
\begin{eqnarray}
\mathrm{MSE} = \frac{\sum 
\left( y_{\mathrm{true}} - y_{\mathrm{pred}}\right)^2}{n}
\, ,
\end{eqnarray}
and $\beta$ acts as a scaling factor between different 
units and is set to 100.
Fig.~\ref{fig:training} shows the learning curve of four training sets, 
and we can observe that the learning procedure reaches a global minimum 
for the loss function.
This behavior demonstrates that our NN model does not suffer from bias 
problems. 
Also, the fact that the validation set follows the same learning curve 
indicates that the data are not overfitted, which means that the 
neural-network parameters have been correctly selected, avoiding variance 
problems.

\vspace{0.01\textheight}

\subsection{Features}
\label{sec:features}

Another aspect that must be taken into account when designing the 
NN model concerns the features to be considered from 
the input data. 
The network was trained with various sets of input features, and we 
concluded that it is possible to accurately determine 
the electronic stopping power of atoms with only five 
input features:
the atomic mass of the incident ion, the
atomic mass of the target, the atomic numbers of both the projectile and 
the target, and the incident projectile energy (per unit mass, i.e., keV/amu).
The corresponding MAPE values are summarized in Table \ref{table:features}.
The first improvement is introduced by considering the logarithm of the 
incident energy in place of the actual value. 
We also added the first ionization energy of the target, obtaining 
a further improvement in the MAPE values. Encouraged by this, we 
tested adding other features, such as the second ionization limit of the 
target, the ionization energy of the ion, and the electronegativities of 
both of them, resulting in poorer outcomes with higher error values.

\begin{table}[H]
\begin{center}
\begin{tabular}{| l | c |}
\hline
\multicolumn{1}{|c|}{Features}  &   MAPE ($\%$) \\
\hline
Default: $Z_p$, $m_p$, $Z_t$, $m_t$, $E$         &              5.76 \\
$E \longrightarrow \log E$            &              5.47 \\
+ first ionization (target)             &    {\bf 5.07} \\
+ first + second ionization (target)    &              14.9 \\
+ first ionization (target) + first ionization (projectile) &        16.1 \\
+ first ionization and electronegativity (target)  & 5.11 \\
+ first ionization (target) + electronegativity (projectile) & 23.8 \\
\hline
\end{tabular}
\end{center}
\caption{Error values evaluated on the cross-validation  dataset, 
resulting with different input features considered in the model training. 
Boldface denotes the selected input features used in the model.}
\label{table:features}
\end{table}

\section{Results}
\label{sec:results}

This section presents and examines the results obtained with the 
{\sc espnn} model, which has been described in Section~\ref{subsec:model}.
The stopping power cross section predictions result from the input 
forward propagation along the neural-network. In the subplots at the 
right of Figs.~\ref{fig:filtradosAu}, \ref{fig:filtradosH}, \ref{fig:filtradosHe}, 
and \ref{fig:filtradospred}, we show the {\sc espnn} predictions for the
collisional systems examined in Section~\ref{subsec:filter}. 
For these ion--target combinations, the model replicates very well the 
experimental results. 
It is noteworthy that although the predictions have been obtained 
individually for each projectile energy value, the model's results are 
given by smooth curves. 
The stopping power cross sections shown in Fig.~\ref{fig:filtradosAu} 
for multiple collisional systems with Au as the target tend to follow 
the trend given by the latest observations. On the contrary, the 
{\sc espnn} results illustrated in Fig.~\ref{fig:filtradosH} for the 
collisional systems with H as projectiles replicate the shape of the 
more statistically representing data. 
The smoothness of the {\sc espnn} stopping power cross sections can be 
understood as a consequence of using a reduced number of features in the 
training process, which also avoids overfitting and any renormalization 
of the input variables. Nonetheless, having a result of a soft curve is 
somewhat remarkable since the experimental input data may present 
significant spreading in some energy regions. 

\begin{figure}[H]
            \centering
	     \includegraphics[width=0.55\textwidth]{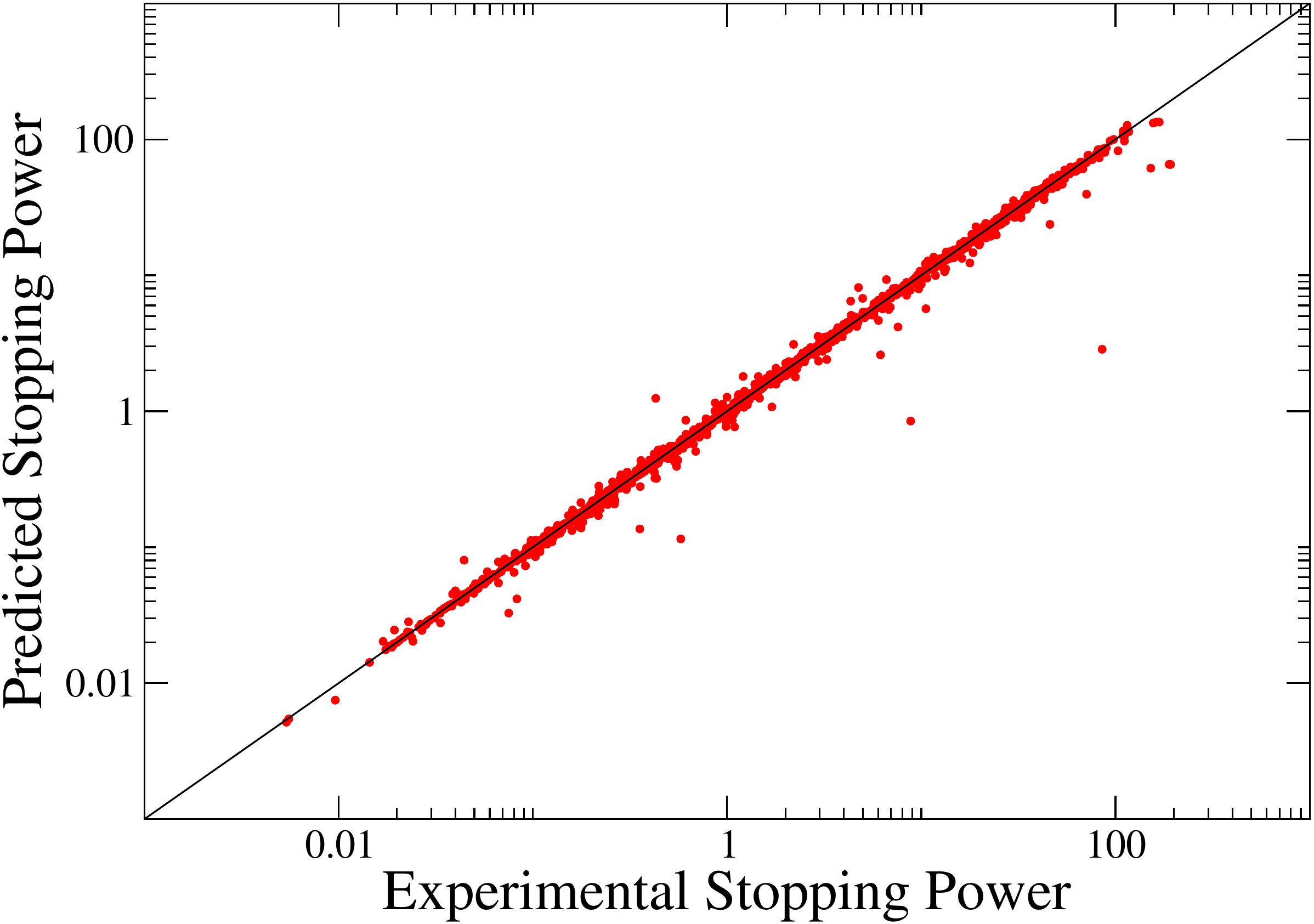} 
            \caption{ Predicted {\sc espnn} values for the  
                       stopping power cross sections, compared with the 
                       experimental results. All the results correspond 
                       to the test dataset, never seen before by the 
                       model.}
        \label{fig:expvspred} 
        \end{figure}

To verify the transferability of the model, we calculated the stopping 
power cross sections for the experimental data points initially excluded 
from the training procedure: the test set. Even though these values were 
not included in the training (or validation) sets, our model can 
reproduce these results with high accuracy, as shown in 
Fig.~\ref{fig:expvspred}.

        \vspace{0.01\textheight}

\begin{figure}[H]
            \centering
	     \includegraphics[width=0.4\textwidth]{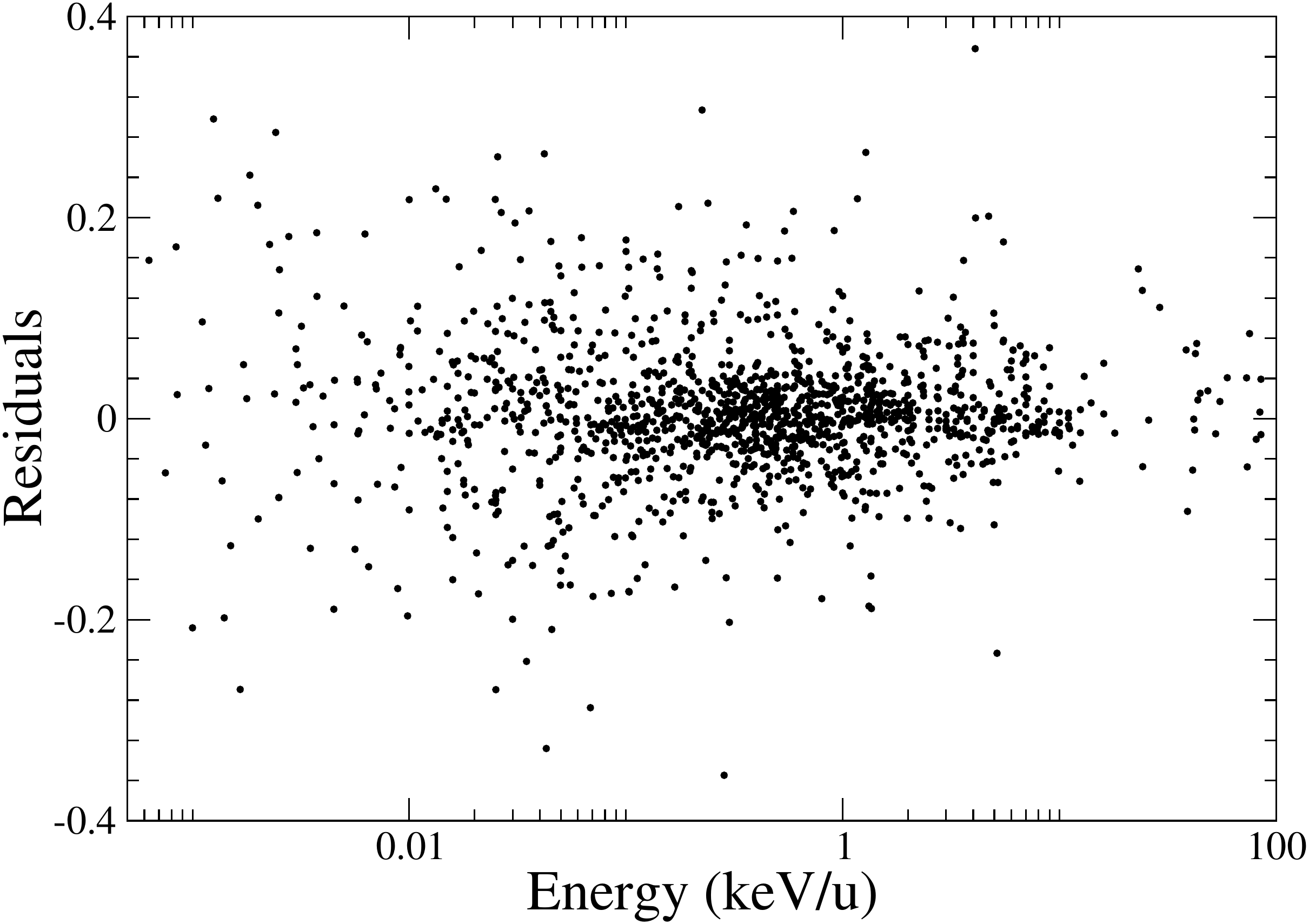} 
	     \includegraphics[width=0.4\textwidth]{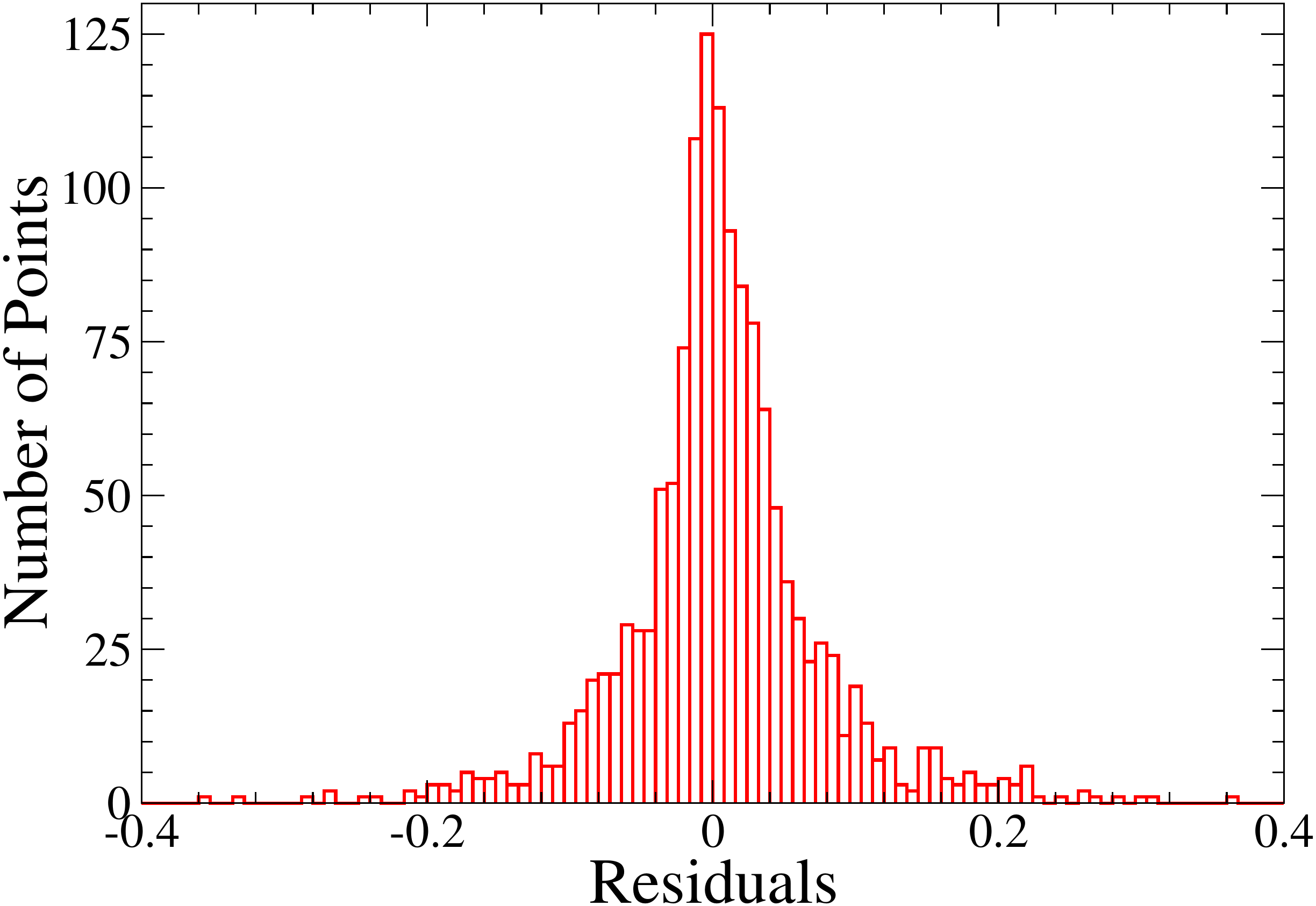} 
            \caption{ Error of the predicted {\sc espnn} values for the 
            test dataset. Left: residual values. Right: Histogram of 
            the prediction errors.}
        \label{fig:histo} 
        \end{figure}

We can further examine the performance of the {\sc espnn} on the test 
set by analyzing the residuals given by 
\begin{eqnarray}
R \equiv 
\frac{ y_{\mathrm{true}} - y_{\mathrm{pred}} }{y_{\mathrm{true}} } \, 
\nonumber \, .
\label{eq:residuals}
\end{eqnarray}
In Fig.~\ref{fig:histo}, we show the test set residuals over the 
experimental energy range (left panel) and the frequency distribution of 
these values (right panel).
We can estimate a mean error value of 5\% from this figure. Because the 
experimental data are scattered over a broad range of uncertainties and 
discrepancies, the present model's accuracy can be considered exceptional.

       \vspace{0.01\textheight}

For a quantitative comparison of our model with other approaches, we 
show in Table~\ref{table:RDDLvsESPNN} the  
MAPE values reported by {\sc srim}\cite{Ziegler:10}, 
the random forest (RF) algorithm from 
Parfitt and Jackman \cite{Parfitt:20}, the deep learning 
(DL) algorithm from Guo {\it et al.} \cite{Guo:22}, and 
the present {\sc espnn} calculations. 
The table does not provide a rigorous comparison between these
methods because it displays the values as reported by the 
respective authors, and different numbers of data points were considered 
in the final MAPE evaluation. 
For that reason, we separated the table into two parts, 
showing on the left the published errors for the training set
and, on the right, the MAPE in the test set. However, we can only make a 
proper comparison with the RF calculations 
reported by Parfitt and Jackman. 
Although the size and characteristics of this work's input and test sets 
were chosen differently than ours, these authors addressed the 
training and an unseen test set separately. 
The MAPE values in Table~\ref{table:RDDLvsESPNN} provide a clear 
picture of the various code's performance. Furthermore, the {\sc espnn}
results on the unseen test set are excellent. A few details are worth 
mentioning; {\sc srim} is a semiempirical code that considers all the 
experimental data published until 2013; therefore, the respective errors 
do not necessarily correspond to its predictive capability on later 
measurements. Inversely, RF describes the training data very well but 
overfits; the high MAPE values for the test set may be an indication of 
this. 

\begin{table}[H]
\begin{center}
\begin{tabular}{|c||c|c|c||c|c|}
\hline
\multirow{2}{*}{Ions} & \multicolumn{3}{c||}{Train} & \multicolumn{2}{c|}{Test}   \\ 
\cline{2-6}
&\ \  SRIM \ \ &\ \   RF \ \   & 
\ \  DL \ \   &\ \   RF \ \   &\ \   ESPNN \ \   \\ 
\hline
H    & 4.0  & 2.4 & 5.6 & 12 & {\bf 4.5}   \\  
He   & 3.9  & 2.1 & 6.0 & 9.1  & {\bf 4.1}  \\  
Li   & 4.8  & 2.1 & 4.2 & 25 & {\bf 7.7} \\  
Be-U & 5.8  & 2.0 & 7.0 & 12  & {\bf 7.1}  \\  
All  & 4.6  & 2.1 & 5.7 & 23 & {\bf 5.7}  \\ 
\hline
\end{tabular}
\end{center}
\caption{Comparisons between the {\sc mape} values 
reported by 
{\sc srim}\cite{Ziegler:10}, the random forest (RF) algorithm from 
Parfitt and Jackman \cite{Parfitt:20}, the deep learning 
(DL) algorithm from Guo {\it et al.} \cite{Guo:22}, and 
the present calculations ({\sc espnn}). 
Left columns: error values in the training set. 
Right columns: error values in the test set. }
\label{table:RDDLvsESPNN}
\end{table}

Because of the importance and the widespread usage of {\sc srim}, we
devised a detailed experiment to compare our model's calculations. 
Attempting to make a fair comparison between our predictions and 
{\sc srim}, we conducted the following procedure: we trained the 
neural-network only with results published before 2013. 
This means that we applied our {\sc dbscan}-based heuristic to all the 
data published up to 2013. 
The remaining experimental values (published after 2013) were set aside, 
and left unclean to avoid data leakage. 
The excluded set (the new test set), was further separated 
into different groups in an iterative fashion. 
First, we considered all the experimental results published after 2013, 
and next, all the publications that appeared after 2015, and successively 
reduced the test set to include only the most recent publications.
Noteworthily, this neural-network was trained only once, with the data 
collected in the IAEA database until 2013. 
In this way, we can compare the prediction power of both methods on an 
equal footing. 
The results are summarized in Table~\ref{table:SRIMvsESPNN}.
This table shows that the {\sc espnn} prediction power is consistently
better than {\sc srim}'s, in all the cases. 
Also, comparing these results with the MAPE values obtained with the 
complete data training, we observe that the prediction power is improved, 
as expected, as new values are included in the data set.

\begin{table}[H]
\begin{center}
\begin{tabular}{ccccccccc}
\hline
\multicolumn{1}{|c|}{}     & \multicolumn{2}{c|}{\textgreater{}2013}                & \multicolumn{2}{c|}{\textgreater{}2015}                & \multicolumn{2}{c|}{\textgreater{}2017}                & \multicolumn{2}{c|}{\textgreater{}2019}                \\ \hline
\multicolumn{1}{|c|}{Ions} & \multicolumn{1}{c|}{{\sc espnn}} & \multicolumn{1}{c|}{{\sc srim}} & \multicolumn{1}{c|}{{\sc espnn}} & \multicolumn{1}{c|}{{\sc srim}} & \multicolumn{1}{c|}{{\sc espnn}} & \multicolumn{1}{c|}{{\sc srim}} & \multicolumn{1}{c|}{{\sc espnn}} & \multicolumn{1}{c|}{{\sc srim}} \\ \hline
\multicolumn{1}{|c|}{H}    & \multicolumn{1}{c|}{7.0}   & \multicolumn{1}{c|}{19.2} & \multicolumn{1}{c|}{4.6}   & \multicolumn{1}{c|}{15.3} & \multicolumn{1}{c|}{6.6}   & \multicolumn{1}{c|}{13.3} & \multicolumn{1}{c|}{3.5}   & \multicolumn{1}{c|}{7.1}  \\ \hline
\multicolumn{1}{|c|}{He}   & \multicolumn{1}{c|}{8.4}   & \multicolumn{1}{c|}{10.6} & \multicolumn{1}{c|}{9.3}   & \multicolumn{1}{c|}{10.1} & \multicolumn{1}{c|}{9.3}   & \multicolumn{1}{c|}{10.1} & \multicolumn{1}{c|}{5.5}   & \multicolumn{1}{c|}{8.3}  \\ \hline
\multicolumn{1}{|c|}{Be-U} & \multicolumn{1}{c|}{6.2}   & \multicolumn{1}{c|}{6.6}  & \multicolumn{1}{c|}{5.4}   & \multicolumn{1}{c|}{6.5}  & \multicolumn{1}{c|}{5.3}   & \multicolumn{1}{c|}{6.3}  & \multicolumn{1}{c|}{}   & \multicolumn{1}{c|}{}     \\ \hline
\multicolumn{1}{|c|}{all}  & \multicolumn{1}{c|}{7.0}   & \multicolumn{1}{c|}{11.4} & \multicolumn{1}{c|}{6.0}   & \multicolumn{1}{c|}{9.8}  & \multicolumn{1}{c|}{6.8}   & \multicolumn{1}{c|}{8.9}  & \multicolumn{1}{c|}{4.1}   & \multicolumn{1}{c|}{7.4}  \\ 
\hline
\hline
\end{tabular}
\end{center}
\caption{Predictive {\sc mape} of the semi-empirical  
{\sc srim} code \cite{Ziegler:10} and the neural-network {\sc espnn} model 
trained only with data collected before 2013.
The first columns show the error predictions for all the 
experimental values reported after 2013. 
The second includes all the values reported after 2015. 
The last columns show the error of the predictions for the 
experimental results produced after 2019.}
\label{table:SRIMvsESPNN}
\end{table}

\section{Conclusions}
\label{sec:conclusions}

In this work, the difficult task of predicting the electronic stopping 
power cross sections for atomic targets is addressed using two ML 
methods. First, a clustering-based algorithm was developed to eliminate 
automatically suspicious or erroneous experimental values. The method 
implements the {\sc dbscan} algorithm to detect outliers and clusters 
with similar data. A multi-step cleaning algorithm analyses these results 
and implements three specific criteria. The algorithm was prepared to 
filter about 20\% of the 36000 experimental atomic results in the IAEA 
stopping power database.

The filtered data were used to train a deep neural-network, constituting 
the second ML method. The neural-network was trained to accurately 
reproduce the training data, producing MAPE results of less than 6\%. 
Considering the widespread of the experimental values, it represents  
excellent performance. The same MAPE was obtained for the test set, 
which constituted data not included in the training procedure. These 
results demonstrate the excellent prediction power of our model.

Moreover, an open-access {\sc espnn} code with the forward propagation 
model has been published in a public repository and is now available. 
This algorithm allows computing the stopping power cross section of any 
collisional system effortlessly and quickly. Certainly, the model will 
be improved in the future with the addition of newer experimental data 
and stopping cross sections of complex molecules.

\begin{acknowledgments}
The following institutions of Argentina financially support this research: 
the CONICET by the PIP11220200102421CO, 
the ANPCyT, PICT-2020-SERIE A-01931, and the
University of Buenos Aires by the project 20020170100727BA. 
\end{acknowledgments}

\section*{Data Availability Statement}
The data that support the findings of this work are openly available in 
GitHub at \href{https://github.com/ale-mendez/ESPNN}{https://github.com/ale-mendez/ESPNN}\cite{github}
and in PyPI at \href{https://pypi.org/project/ESPNN/}{https://pypi.org/project/ESPNN/}\cite{pypi}.

\newpage
\begin{appendix}
\section{Machine Learning Filtering Procedure}
\label{app:filtering}

\begin{algorithm}[H]
\setstretch{1.1} 
\begin{algorithmic}[1]
\Procedure{ Clusters and Outliers}{}
   \State {\bf Input}:
	\begin{enumerate}
           \item  $\epsilon$ (radius of the neighborhood)
           \item $N_{\mathrm{min}}$ (number of reachable points)
        \end{enumerate}
   \State Run {\sc dbscan} to select different clusters and outliers.
   \State {\bf Output}:
	\begin{enumerate}
           \item  $N_{\mathrm{clust}}$ (number of clusters)
           \item  $N_{\mathrm{outl}}$ (number of noise points)
        \end{enumerate}
\EndProcedure{ {\sc clusters and outliers}   }{}
\\
\Procedure{Drop Publications}{}
\For{ list of publications $P_i$}

\\

\Procedure{Check Newer Publications}{}
\State Count $n_i$ (number of experimental results $P_i$)
\State Define $\Delta E_i$ (energy range covered in $P_i$).
\For{ list of publications $P_j^n {\bf ~newer~than~} P_i$ }
 \State Define $\Delta E_{ji}^n \equiv \Delta E_j^n \bigcap \Delta E_i$.
\EndFor
 \State Define $\Delta E^{\mathrm{new}} \equiv \bigcup\limits_{j} \Delta E_{ji}^n$
\State {\bf If} $\frac{\Delta E^{\mathrm{new}}}{\Delta E_i} \leq \sigma_{\Delta}$: 
          {\bf KEEP} publication $P_i$ and {\bf BREAK}
\EndProcedure{ {\sc check newer publications}  }{}
   
\\
   
\Procedure{Isolated Results}{}
       \State Count $ N^i_{\mathrm{outl}}$ (number of outliers in $P_i$) 
       \State {\bf If} $\frac{N^i_{\mathrm{out}} }{n_i} > \sigma_{\mathrm{outl}} $:  
        {\bf DROP} publication $P_i$
\EndProcedure{ {\sc isolated results}  }{}

\\
   
\Procedure{Test Cluster}{}
       \State Identify $C^i$ (biggest cluster in $P_i$)
       \State Count $l^i$ (number of $n_i$ values from $P_i$ $\in C^i$ )
       \State Count $t^i$ (total number of results $\sum n_j \in C^i$ )
       \State {\bf If} $\frac{l^i}{t^i} > \sigma_{\mathrm{clu}} $:  
          {\bf DROP} publication $P_i$
\EndProcedure{ {\sc test clusters}  }{}

\\

\EndFor

\EndProcedure{ {\sc drop publications}  }{}
\end{algorithmic}
\end{algorithm}

\section{Instructions for running {\sc espnn} }
\label{app:code}

The code is public and can be used remotely or locally. 
Detailed instructions can be found at:
\begin{verbatim}
    https://pypi.org/project/ESPNN/
\end{verbatim}

For remote use, the {\sc espnn} code can run on the Google Colab platform 
(see the url address in the PyPI page).
For a local installation, the code can be installed via pip:
\begin{verbatim}
          pip install ESPNN
\end{verbatim}
or, by downloading or cloning the repository at:
\begin{verbatim}
    https://github.com/ale-mendez/ESPNN
\end{verbatim}

Once the code is installed, the execution is extremely simple. 
For example, if the He (projectile) and Au (target) 
collision is required, the user can type from the terminal 

\begin{verbatim}
          python -m ESPNN He Au
\end{verbatim}
 
\begin{figure}[H]
            \centering
	     \includegraphics[width=0.7\textwidth]{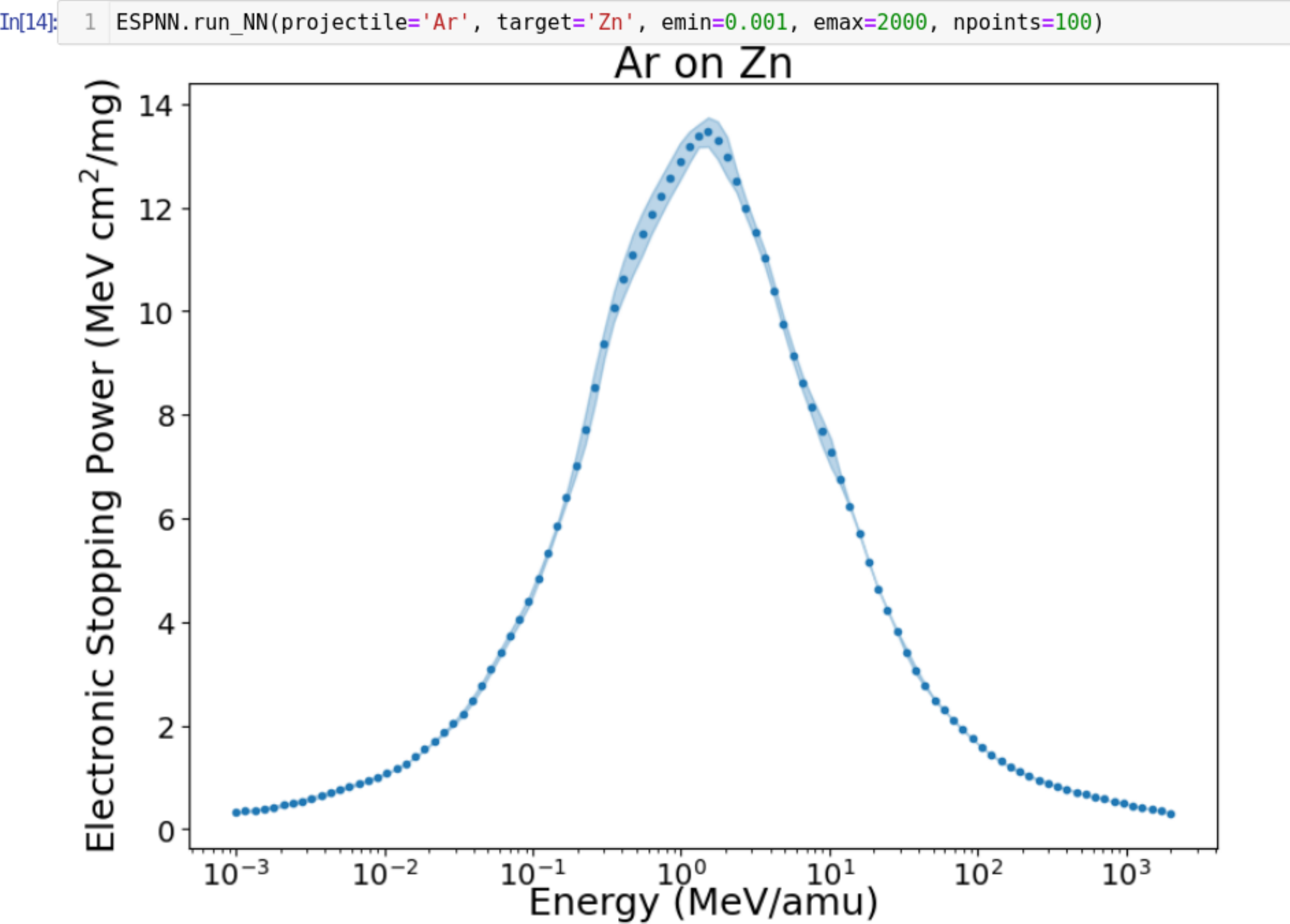} 
             \caption{ Example of an output figure: Ar in Zn. The stopping 
             power cross sections are predicted at npoints=100 points, 
             and the estimated variance is plotted in light blue.}
        \label{fig:captureres} 
        \end{figure}

To run the code using a Jupyter Notebook, only these two 
lines are required: 

\begin{verbatim}
          import ESPNN
          ESPNN.run_NN(projectile="He", target="Au")
\end{verbatim}

There are other optional arguments that can be modified, such as the 
minimum and maximum energy values, the number of energy points in 
the grid, the path to the output file, etc. 
For information about the options:

\begin{verbatim}
          python -m ESPNN -h
\end{verbatim}

As an example, we show the results for Ar in Zn. An output file called 
ArZn\_prediction.dat is generated. This file has three columns:
Energy (MeV/amu), Stopping power (MeV $\mathrm{cm}^2/\mathrm{mg}$), 
and variance (MeV $\mathrm{cm}^2/\mathrm{mg}$).
Figure~\ref{fig:captureres} is generated on the notebook.

       \vspace{0.01\textheight}

\end{appendix}

\providecommand{\noopsort}[1]{}\providecommand{\singleletter}[1]{#1}%


\end{document}